\documentclass[aip,jap,twocolumn,groupedaddress,showpacs,floatfix]{revtex4-1}
\usepackage{amsmath}
\usepackage{amssymb}
\usepackage{graphicx}
\usepackage{colortbl}

\begin{document}

\title{Time-dependent quantum transport and power-law decay of the transient 
current in a nano-relay and nano-oscillator}

\author{Eduardo C. Cuansing}
\email[]{eduardo.cuansing@gmail.com}
\author{Gengchiau Liang}
\email[]{elelg@nus.edu.sg}
\affiliation{Department of Electrical and Computer Engineering, National
University of Singapore, Singapore 117576, Republic of Singapore}

\date{August 11, 2011}

\begin{abstract}

Time-dependent nonequilibrium Green's functions are used to study electron 
transport properties in a device consisting of two linear chain leads and a 
time-dependent interleads coupling that is switched on non-adiabatically. We 
derive a numerically exact expression for the particle current and examine 
its characteristics as it evolves in time from the transient regime to the 
long-time steady-state regime. We find that just after switch-on the current 
initially overshoots the expected long-time steady-state value, oscillates 
and decays as a power law, and eventually settles to a steady-state value 
consistent with the value calculated using the Landauer formula. The power-law 
parameters depend on the values of the applied bias voltage, the strength of 
the couplings, and the speed of the switch-on. In particular, the oscillating 
transient current decays away longer for lower bias voltages. Furthermore, the 
power-law decay nature of the current suggests an equivalent series 
resistor-inductor-capacitor circuit wherein all of the components have 
time-dependent properties. Such dynamical resistive, inductive, and capacitive 
influences are generic in nano-circuites where dynamical switches are
incorporated. We also examine the characteristics of the dynamical current in 
a nano-oscillator modeled by introducing a sinusoidally modulated interleads 
coupling between the two leads. We find that the current does not strictly 
follow the sinusoidal form of the coupling. In particular, the maximum current 
does not occur during times when the leads are exactly aligned. Instead, the
times when the maximum current occurs depend on the values of the bias
potential, nearest-neighbor coupling, and the interleads coupling.

\end{abstract}

% 73.63.-b: Electronic transport in nanoscale materials and structures
% 72.10.Bg: General formulation of transport theory
% 73.23.-b: Electronic transport in mesoscopic systems

\maketitle

%----------------------------------------------------------------------------
%                        Introduction
%----------------------------------------------------------------------------
\section{INTRODUCTION}
\label{sec:intro}

The further miniaturization of electronic devices will eventually lead to 
molecular electronics wherein particles pass through molecular-scale devices 
whose constituent molecules may have been manipulated and synthetically 
assembled or created.\cite{heath03} Molecular transistors, in particular, 
are of significant practical interests and whose successful implementations 
are currently actively being pursued. Several theoretical models of the 
transistor have been proposed\cite{palacios02,ghosh04} and experimental 
successes have also been reported.\cite{kubatkin03} A related molecular-scale 
device, the molecular switch, has also garnered significant interests because 
of the switch's important role in circuit design and architecture. A 
theoretical model of the switch makes use of a mechanism that involves the 
reversible displacement of an atom in a molecular wire through the application 
of a gate voltage.\cite{wada93} In addition, experimental realizations of the 
atomic switch include mechanisms involving the reversible transfer of an atom 
between two leads,\cite{eigler91} the dynamical onset of single-atom contact 
between leads\cite{terabe05} and the manipulation of atomic bonds using a 
dynamic force microscope.\cite{sweetman11} Having a dynamical switch in a 
nano-circuit, however, necessitates the appearance of time-dependent behavior, 
particularly during the transient regime just after switch-on. The circuit 
transitions from being disconnected into connected in a short time and the 
current does not instantly switches into the steady-state value upon 
connection. It is therefore informative to know the characteristics of the 
current just after switch-on and during the transient regime, and determine 
how this current approaches the steady-state value. In this work, we introduce 
a model device representing a system wherein the current can be dynamically 
toggled on and off. There are several theoretical approaches in treating 
time-dependent quantum transport. Among these include time-dependent
density functional theory,\cite{runge84} propagating the Kadanoff-Baym
equations,\cite{prociuk08} Floquet theory,\cite{moskalets02} path-integral
techniques,\cite{altland10} and the density matrix renormalization group
method.\cite{branschadel10,karrasch10,pletyukhov10} In this work we choose
to use time-dependent nonequilibrium Green's functions (TD-NEGF) because, as 
is shown in this paper, dynamically toggling the coupling between the leads 
is straightforward using the technique, even without the assumption of weak 
coupling, and the resulting expression for the time-dependent current is 
numerically exact. 

\begin{figure}[h!]
\includegraphics[width=3in,clip]{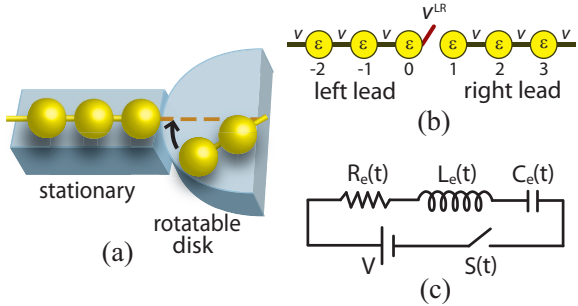}
\caption{(Color online) (a) An illustration of the nano-relay and 
nano-oscillator. The left lead is firmly on a stationary substrate and the 
right lead is on a rotatable disk. The left and right leads align when the 
disk is rotated clockwise to the dash line. (b) The device can be represented 
by two semi-infinite leads connected by a time-varying coupling 
$v^{\rm LR}(t)$ switched on at $t = 0$. In both leads, the on-site energy is 
$\epsilon$ and the nearest-neighbor hopping parameter is $v$. (c) An 
equivalent resistor-inductor-capacitor circuit with dynamical properties.
\label{fig:main}}
\end{figure}

The device we examine consists of a semi-infinite linear chain of atoms, 
i.e., the left lead, which is stationary and another semi-infinite linear 
chain of atoms, i.e., the right lead, that is on a rotatable disk. An
illustration of the device is shown in Fig.~\ref{fig:main}(a). When the disk 
is rotated clockwise to the dashed line, the left and right leads align and 
conduction occurs. We model how the two leads couple in two ways: as an 
abrupt Heaviside step function and as a hyperbolic tangent that gradually 
progresses in time. For such cases, the off state has no current flowing
across the leads. In addition, we can also model an oscillator by swinging 
the disk back and forth across the dashed line. The rotatable disk can also 
be replaced by a gate voltage, located on top of the right lead, that could 
rotate the right lead to its desired position. This latter configuration has 
previously been examined and, in particular, the steady-state transport 
properties of its on and off states have been studied \cite{ghosh04}. In this 
paper, however, we study the full time-dependent transport properties of the 
device. Numerically exact expressions for the current and the needed 
nonequilibrium Green's functions are derived. We then show that just after 
switch-on and during the transient regime the current initially overshoots 
the expected value of the steady-state current and then oscillates around 
the steady-state value while decaying as a power law. Such a power-law 
decaying transient current suggests the appearance of dynamical resistance, 
inductance, and capacitance in the system during the transient regime. 
Power-law decaying currents have also recently been predicted to appear in
a system containing a quantum dot channel\cite{karrasch10} and in the 
anisotropic Kondo model.\cite{pletyukhov10}

%---------------------------------------------------------------------------
%                        Model and method
%---------------------------------------------------------------------------
\section{MODEL AND METHOD}
\label{sec:model}

We implement time-dependent nonequilibrium Green's functions techniques to 
investigate the dynamical transport properties of particles traversing through 
a device with time-varying components. To determine the dynamical transport 
properties, a TD-NEGF approach can be used that utilizes either
two-time\cite{jauho94,haug07} Green's functions, or 
double-energy\cite{anantram95} Green's functions, or Green's functions that
depend on one time and one energy variables.\cite{arrachea05a,arrachea05b} In 
a transistor with source-channel-drain and top gate configuration, for 
example, the device can be modeled by a Hamiltonian constructed using density 
functional theory\cite{zhu05,ke10} or tight-binding theory\cite{moldoveanu07} 
and the dynamical transport properties of small channels are calculated using 
TD-NEGF. For devices with larger channels, a self-consistent calculation 
based on the Poisson equation and TD-NEGF can be done to determine the 
consistent dynamical potential and charge density in the 
channel.\cite{kienle09} In this work, in contrast, we study a device 
consisting only of two leads, i.e., there is no channel between the leads. 
The time-dependence comes from non-adiabatically toggling on the coupling 
between the leads. We use TD-NEGF to derive a numerically exact expression 
for the dynamical current and investigate how the devices we call a 
nano-relay and a nano-oscillator respond to time-varying influences. This 
approach has recently also been used in the study of dynamical heat transport 
in a thermal switch\cite{cuansing10}.

We model the system by the total Hamiltonian
$H = H^{\rm L} + H^{\rm R} + H^{\rm LR}$, where $H^{\rm L}$ is for the 
left lead on the stationary substrate, $H^{\rm R}$ is for the right lead 
on the rotatable disk, and $H^{\rm LR}$ includes the dynamic interleads 
coupling. In the leads, particles follow the tight-binding Hamiltonian
\begin{equation}
H^{\alpha} = \sum_k \epsilon_k^{\alpha}~c_k^{\alpha \dagger} c_k^{\alpha} +
\sum_{kj} v_{kj}^{\alpha}~c_k^{\alpha \dagger} c_j^{\alpha},~~
\alpha = {\rm L, R},
\label{eq:leadhamiltonian}
\end{equation}
where $c_k^{\alpha \dagger}$ and $c_k^{\alpha}$ are particle creation and
annihilation operators at the $k$th site in the $\alpha$ lead.
$\epsilon_k$ is the on-site energy at site $k$ and $v_{kj}$ is the 
hopping parameter between nearest-neighbor sites $k$ and $j$ (see
Fig.~\ref{fig:main}(b)). The sums are over all sites in the $\alpha$ lead. 
$H^{\rm LR}$ includes the time-dependent component of the total Hamiltonian 
and is of the form
\begin{equation}
H^{\rm LR} = \sum_{kj} \left( v_{kj}^{\rm LR}~c_k^{{\rm L} \dagger}
c_j^{\rm R} + v_{jk}^{\rm RL}~c_j^{{\rm R} \dagger} c_k^{\rm L} \right),
\label{eq:HLR}
\end{equation}
where $v_{jk}^{\rm LR}(t) = v_{kj}^{\rm RL}(t)$ is the time-dependent 
coupling between the left and right leads and is switched-on at $t = 0$. 
Only the right-most site of the left lead, i.e., the site labeled $0$
in Fig.~\ref{fig:main}(b), can couple to the left-most site of the right 
lead, i.e., the site labeled $1$ in the figure.

The current can be determined by noting how the number operator, 
$N^{\alpha} = \sum_k c_k^{\alpha \dagger} c_k^{\alpha}$, changes with time,
i.e., $I^{\rm R}(t) = -q \left< d N^{\rm R}/dt \right>$, where $q$ is the 
electron charge. Defining the two-time lesser Green's function as
\begin{equation}
G_{jk}^{{\rm RL},<}(t_1,t_2) = \frac{i}{\hbar} 
\left<c_k^{{\rm L} \dagger}(t_2)~c_j^{\rm R}(t_1)\right>,
\label{eq:lesserGF}
\end{equation}
we can write the electron current flowing out of the right lead as
\begin{widetext}
\begin{equation}
I^{\rm R}(t) = -\frac{iq}{\hbar} \sum_{kj}\left(v_{kj}^{\rm LR}
\left<c_k^{{\rm L}\dagger} c_j^{\rm R}\right> - v_{jk}^{\rm RL}
\left<c_j^{{\rm R}\dagger} c_k^{\rm L}\right>\right)
= -2q {\rm Re}\left(\sum_{kj} v_{kj}^{\rm LR}~G_{jk}^{{\rm RL},<}(t,t)\right).
\label{eq:rightcurrent}
\end{equation}
\end{widetext}
Similar steps can be done to determine the current flowing out of the left 
lead. We find $I^{\rm L}(t) = - I^{\rm R}(t)$ and therefore, current is 
always conserved at each instant of time $t$.

We define the contour-ordered Green's function
\begin{equation}
G_{jk}^{\rm RL}(\tau_1,\tau_2) = -\frac{i}{\hbar} \left< {\rm T}_c~
c_j^{\rm R}(\tau_1) c_k^{{\rm L} \dagger}(\tau_2)\right>,
\label{eq:contourGF}
\end{equation}
where T$_c$ is the contour-ordering operator and $\tau_1$ and $\tau_2$ are
variables along the contour.\cite{haug07} Since we want to calculate the 
current for both the steady-state and non-steady-state regimes, we employ 
a contour that begins at $t = 0$ when the interleads coupling has just been
switched on, then proceeds to time $t$ where we want to determine the 
current, and then goes back to time $t = 0$. In the interaction picture the 
contour-ordered Green's function can be expanded in powers of $i/{\hbar}$. 
Applying Wick's theorem to the resulting expansion and then using Langreth's 
theorem and analytic continuation,\cite{haug07} we get a numerically
exact expression that includes all terms in the expansion for the lesser
Green's function:
\begin{widetext}
\begin{equation}
\begin{split}
G_{jk}^{{\rm RL},<}(t_1,t_2) = {} & G_{jk,1}^{{\rm RL},<}(t_1,t_2)
  +~\sum_{mn} \int_0^t dt'~G_{jm}^{{\rm RL},r}(t_1,t')
  v_{mn}^{\rm LR}(t') G_{nk,1}^{{\rm RL},<}(t',t_2) \\
  & +~\sum_{mn} \int_0^t dt'~G_{jm,1}^{{\rm RL},<}(t_1,t')
  v_{mn}^{\rm LR}(t') G_{nk}^{{\rm RL},a}(t',t_2) \\
  & +~\sum_{mnpq} \int_0^t dt' \int_0^t dt''~
  G_{jm}^{{\rm RL},r}(t_1,t') v_{mn}^{\rm LR}(t')
  G_{np,1}^{{\rm RL},<}(t',t'') v_{pq}^{\rm LR}(t'') 
  G_{qk}^{{\rm RL},a}(t'',t_2),
\end{split}
\label{eq:gless}
\end{equation}
where the advanced and retarded Green's functions are
\begin{equation}
G_{jk}^{{\rm RL},\zeta}(t_1,t_2) = G_{jk,1}^{{\rm RL},\zeta}(t_1,t_2)
  +~\sum_{mn} \int_0^t dt'~G_{jm,1}^{{\rm RL},\zeta}(t_1,t')
  v_{mn}^{\rm LR}(t') G_{nk}^{{\rm RL},\zeta}(t',t_2),~~\zeta = r,a.
\label{eq:advret}
\end{equation}
The first-order retarded and advanced Green's functions are
\begin{equation}
G_{jk,1}^{{\rm RL},\zeta}(t_1,t_2) = \int_0^t dt'
  ~g_{jj}^{{\rm R},\zeta}(t_1-t') v_{jk}^{\rm RL}(t')
  g_{kk}^{{\rm L},\zeta}(t'-t_2),
\label{eq:advretfirst}
\end{equation}
and the first-order lesser Green's function is
\begin{equation}
G_{jk,1}^{{\rm RL},<}(t_1,t_2) = \int_0^t dt' 
  \left\{g_{jj}^{{\rm R},r}(t_1-t') v_{jk}^{\rm RL}(t') 
  g_{kk}^{{\rm L},<}(t'-t_2)
  + g_{jj}^{{\rm R},<}(t_1-t') v_{jk}^{\rm RL}(t')
  g_{kk}^{{\rm L},a}(t'-t_2)\right\},
\label{eq:lesserfirst}
\end{equation}
\end{widetext}
where the $g^r(t)$, $g^a(t)$, and $g^<(t)$ are the retarded, advanced, and 
lesser free-leads Green's functions, respectively. Time-translation 
invariance is satisfied by the free leads and therefore, their corresponding
Green's functions can be calculated in the energy domain using the techniques 
of steady-state NEGF.\cite{datta05} The integrals in 
Eqs.~(\ref{eq:advretfirst}) and (\ref{eq:lesserfirst}) are then determined 
using the extended Simpson's rule algorithm for numerical 
integration.\cite{press07} Furthermore, the expressions for the advanced and 
retarded Green's functions in Eq.~(\ref{eq:advret}) are in the form of a 
Fredholm equation of the second kind and can be solved by discretizing the 
time integrals, based on the extended Simpson's rule algorithm, and performing 
a matrix inversion using LU (lower-triangular and upper-triangular) 
decomposition.\cite{press07} The lesser Green's function can then be 
calculated from the retarded and advanced Green's function by applying the 
extended Simpson's rule algorithm to numerically solve the integrals in 
Eq.~(\ref{eq:gless}).

%----------------------------------------------------------------------------
%                        Results and discussion
%----------------------------------------------------------------------------
\section{RESULTS AND DISCUSSION}
\label{sec:results}

Firstly, we examine the impact of the switch-on speed to current 
characteristics. The functional form of the interleads coupling, 
$v^{\rm LR}(t)$, describes how the device is switched on. We examine two 
types of switch-ons: an abrupt Heaviside step function switch-on and a 
gradually progressing hyperbolic tangent switch-on of the form 
$v^{\rm LR}(t) = v^{\rm LR} \tanh(\omega_d t)$, where $\omega_d$ is the 
driving frequency. The step function switch-on is actually the limit when 
$\omega_d\rightarrow\infty$ of the hyperbolic tangent switch-on. Furthermore, 
we set the on-site energy $\epsilon = 0$. The left and right leads have 
temperature $T^{\rm L} = T^{\rm R} = 300~{\rm K}$ and chemical potential 
$\mu_{\rm L} = \epsilon_F$ and $\mu_{\rm R} = \epsilon_F - q V_b$, where we 
set the Fermi energy $\epsilon_F = 0$. The bias potential $V_b$ is applied to 
the right lead and the bias potential energy is written as $U_b = q V_b$.

\begin{figure}[h!]
\includegraphics[width=3in,clip]{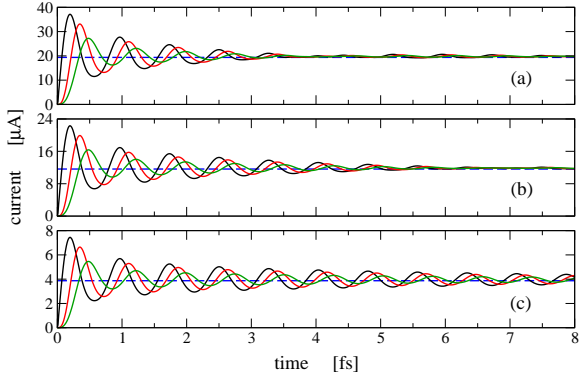}
\caption{(Color online) The current as a function of time when the interleads 
coupling $v^{\rm LR}(t)$ is switched on as a step function at $t = 0$ (black 
lines), gradually as a hyperbolic tangent with driving frequency 
$\omega_d = 0.5~{\rm [1/t]}$ (red lines), and $\omega_d = 0.25~{\rm [1/t]}$ 
(green lines), where ${\rm [1/t]} = 10^{16}~{\rm rad/s}$. The bias 
potentials are ${\rm (a)}~U_b = 0.5~{\rm eV}$, ${\rm (b)}~U_b = 0.3~{\rm eV}$,
and ${\rm (c)}~U_b = 0.1~{\rm eV}$. The (blue) dashed lines are the values of
the steady-state current. The hopping parameter in the leads is 
$v = -2.7~{\rm eV}$.
\label{fig:tanh}}
\end{figure}

Figure \ref{fig:tanh} shows the current flowing out of the left lead for the 
step function and hyperbolic tangent switch-on with driving frequencies 
$\omega_d = 0.25~{\rm [1/t]}$ and $0.5~{\rm [1/t]}$, where
${\rm [1/t]} = 10^{16}~{\rm rad/s}$. The steady-state values of the current 
calculated separately using the Landauer formula for a linear chain with unit 
transmission\cite{datta97} are also shown in Fig.~\ref{fig:tanh} as dashed 
lines. During the times just after the switch-on, the current rapidly 
increases and overshoots the expected long-time steady-state value. It then 
oscillates and decays in time, eventually settling to the steady-state value. 
It can be seen that as the driving frequency is decreased the amplitude of 
the oscillations also decreases. In addition, the peaks are displaced to later 
times because of the more gradual progress of the interleads coupling 
$v^{\rm LR}(t)$. In Fig.~\ref{fig:tanh} we also see the dependence of the 
decay time of the transient current to the applied bias potential and the 
speed of the switch-on. The higher bias results in a faster decay time for 
the oscillating transient current.

The transient current oscillates and decays in time until it settles to a 
steady-state value.\cite{dc} During the transient regime the strength of the 
interleads coupling dynamically changes resulting in particles temporarily 
accumulating at the left and right sides of that coupling. Although we do not 
explicitly consider Coulomb interactions between charges, the temporary 
accumulation of charges at the sides of the interleads coupling, together 
with the distance between the accumulated charges and the applied bias voltage 
across the interleads coupling, can be regarded to generate a temporary 
dynamical capacitance. Similarly, a dynamical inductance may arise because the 
transient current is varying in time. By considering possible equivalent 
circuit combinations and performing least-squares fitting to the envelope of 
the decaying transient current we find that the decay closely follows a power 
law, indicating an equivalent series resistor-inductor-capacitor (RLC) circuit 
whose components have time-dependent properties. In a previous study using 
semiclassical Boltzmann transport theory on quantum wires, it is found that 
the wire can be modeled by an equivalent series RLC circuit.\cite{salahuddin05}
A series RLC circuit consisting of components with constant resistance, 
inductance, and capacitance results in a transient current whose envelope 
decays as an exponential function. However, when the resistance, inductance, 
and capacitance vary in time, the resulting transient current can oscillate 
and could decay as a power law. Making therefore such an analogy to the 
quantum device we are examining (see Fig.~\ref{fig:main}(c)), applying 
Kirchhoff's law to a series RLC circuit with time-varying components leads to 
the equation
\begin{widetext}
\begin{equation}
\frac{d^2 I}{dt^2} + \left(\frac{R}{L}+\frac{1}{L}\frac{dL}{dt}\right)
\frac{dI}{dt} + \left(\frac{1}{L}\frac{dR}{dt} + \frac{1}{LC}\right) I
= \frac{Q}{LC^2}\frac{dC}{dt},
\label{eq:kirchhoff}
\end{equation}
\end{widetext}
where $I$ is the time-dependent current through the circuit, 
$R \equiv R(t)$ is the resistance, $L \equiv L(t)$ is the inductance,
$C \equiv C(t)$ is the capacitance, and $Q=\int_0^t I~dt$ is the 
time-dependent charge accumulating at the capacitor. Furthermore, the
power law fits imply that the current is of the form
\begin{equation}
I(t) = I_0~t^{-\alpha} \sin(\omega t + \phi) + I_0,
\label{eq:powerlaw}
\end{equation}
where $\alpha$ is the power-law exponent determined from the fits, $\omega$ 
is the time-independent frequency of oscillation of the transient current, 
$\phi$ is the phase determined from initial conditions, and $I_0$ is the 
time-independent steady-state current. Taking the time derivative of
Eq.~(\ref{eq:powerlaw}) twice, we find
\begin{equation}
\frac{d^2 I}{dt^2} + \frac{2 \alpha}{t} \frac{dI}{dt} + \omega_t^2 I
= \omega_t^2 I_0,
\label{eq:currenteq}
\end{equation}
where $\omega_t^2 = \omega^2 + \frac{\alpha(\alpha-1)}{t^2}$. Comparing
Eqs.~(\ref{eq:kirchhoff}) and (\ref{eq:currenteq}) we get the coupled 
equations
\begin{equation}
\begin{split}
\frac{R}{L} + \frac{1}{L} \frac{dL}{dt} = {} & \frac{2 \alpha}{t}, \\
\frac{1}{L} \frac{dR}{dt} + \frac{1}{LC} = {} & \omega_t^2, \\
\frac{Q}{LC^2} \frac{dC}{dt} = {} & \omega_t^2 I_0,
\end{split}
\label{eq:identify}
\end{equation}
which can be solved to determine how $R(t)$, $L(t)$, and $C(t)$ vary in time 
for specific values of $\alpha$ and $\omega$.

\begin{figure}[h!]
\includegraphics[width=3in,clip]{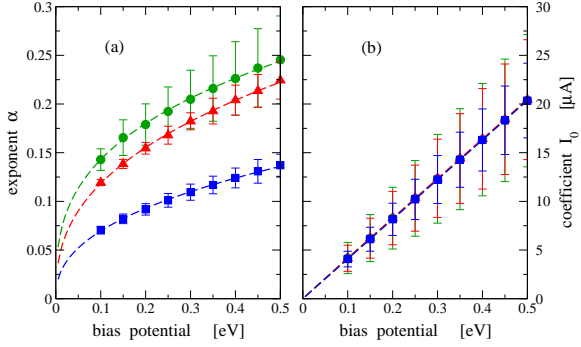}
\caption{(Color online) (a) The power-law exponent $\alpha$ and (b) 
coefficient $I_0$ as functions of the bias potential. The (green) dots are 
for the Heaviside step function interleads coupling. The (red) triangles are 
for the hyperbolic tangent interleads coupling with 
$\omega_d = 0.5~{\rm [1/t]}$. The (blue) squares are for 
$\omega_d = 0.25~{\rm [1/t]}$. The dashed lines are power-law fits to the 
corresponding data points. The amplitude of the couplings are 
$v = v^{\rm LR} = -2.7~{\rm eV}$. Error bars arise from the difference 
between values from the top and bottom envelopes.
\label{fig:decay1}}
\end{figure}

The power-law exponent $\alpha$ determines how fast the transient current
decays until it reaches the steady-state value. In Fig.~\ref{fig:decay1}(a), 
it can be observed that by increasing the bias potential the value of $\alpha$ 
also increases, thereby speeding up the decay of the transient current. 
$\alpha$ and the power-law coefficient $I_0$ actually also follow power-law 
fits when the bias potential is varied, as can be seen in 
Fig.~\ref{fig:decay1}. It suggests $\alpha = \alpha_0~U_b^{\beta}$ and 
$I_0 = I_{00}~U_b^{\gamma}$, where $\alpha_0$ and $I_{00}$ are independent of 
$U_b$. The power-law exponents $\beta$ and $\gamma$ determine how fast 
$\alpha$ and $I_0$, respectively, change when the bias potential is varied. 
In Table~\ref{tbl:tanh} we show how the values of the power-law fitting 
parameters change when $U_b$ is varied.

\begin{table}[h!]
\caption{Values from the power-law fits to $\alpha$ and $I_0$ when the 
interleads coupling is in the form of a hyperbolic tangent and a step 
function. The couplings are $v = -2.7~{\rm eV}$ and 
$v^{\rm LR} = -2.7~{\rm eV}$. The dimension of $\alpha_0$ is 
$(1/{\rm eV})^{\beta}$ and  $I_{00}$ is $\mu {\rm A}/({\rm eV})^{\gamma}$. }
\label{tbl:tanh}
\begin{ruledtabular}
\begin{tabular}{|c|c|c|c|c|}
  \hline
  $\omega_d$ [1/t] & $\alpha_0$ & $\beta$ & $I_{00}$ & $\gamma$ \\
  \hline \hline
  0.25   & 0.181 & 0.416 & 40.753 & 0.999 \\
  0.5    & 0.292 & 0.392 & 40.684 & 0.989 \\
  (step) & 0.307 & 0.332 & 40.214 & 0.983 \\
  \hline
\end{tabular}
\end{ruledtabular}
\end{table}

From Table~\ref{tbl:tanh}, we see that the values of $I_{00}$ and $\gamma$ are 
independent of the speed of the switch-on. In addition, the exponent $\gamma$ 
is about one. These imply that $I_0$ increases linearly with the bias 
potential and is consistent with the identification that $I_0$ is the 
steady-state current. For the exponent $\alpha$, we find that as the speed of 
the switch-on is increased the coefficient $\alpha_0$ also increases but the 
exponent $\beta$ decreases. The increasing $\alpha_0$ suggests that for a 
given bias potential, the faster switch-on results in a faster decay of the 
transient current. Since the slightly decreasing $\beta$ is still positive, 
increasing the bias potential still speeds up the decay of the transient 
current. As a result, when the device is operated under low bias its current 
suffers oscillations and overshootings longer than when it is operated under 
higher bias.  Furthermore, it can be observed that the power-law parameters 
$\alpha_0$ and $\beta$ control the speed of decay of the transient current. 
The values of these two parameters vary depending on the speed of the 
switch-on. If we want the system to have a fast decaying transient current, 
then our results indicate that we need a switch-on that is as fast as 
possible.

\begin{figure}[h!]
\includegraphics[width=3in,clip]{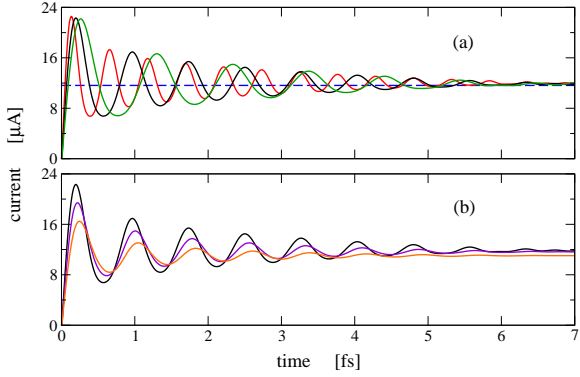}
\caption{(Color online) The current as a function of time for a step function 
switch-on. (a) The couplings have values $v = v^{\rm LR} = -4.0~{\rm eV}$ 
(red line), $v = v^{\rm LR} = -2.7~{\rm eV}$ (black line), and
$v = v^{\rm LR} = -2.0~{\rm eV}$ (green line). The (blue) dashed line is the 
value of the steady-state current. (b) The hopping parameter is fixed at
$v = -2.7~{\rm eV}$ and the interleads coupling is varied with values
$v^{\rm LR} = -2.7~{\rm eV}$ (black line), $v^{\rm LR} = -2.4~{\rm eV}$
(violet line), and $v^{\rm LR} = -2.1~{\rm eV}$ (orange line). The bias 
potential for both plots is $U_b = 0.3~{\rm eV}$.
\label{fig:varyv}}
\end{figure}

Next, we investigate the effects of varying the values of the hopping 
parameter $v$ and the interleads coupling $v^{\rm LR}$. The values of these 
tight-binding parameters depend on the material used and varying them 
effectively means that we change the material we use for the device. The 
parameters can be varied separately or they can be varied while maintaining 
$v = v^{\rm LR}$. Firstly, we consider the latter case. In order to 
determine only the effects of varying the couplings, and not the effects of
the speed of the switch-on, we employ the step function switch-on. The
current as a function of time is shown in Fig.~\ref{fig:varyv}(a).
Since  $v = v^{\rm LR}$, the long-time steady-state values of the current can 
be calculated from the Landauer formula with a transmission coefficient 
$T = 1$ (no scattering involved). This steady-state value is shown in 
Fig.~\ref{fig:varyv}(a) as a dashed line. We examine coupling values 
$v = v^{\rm LR} = -2.0~{\rm eV}, -2.7~{\rm eV}$, and $-4.0~{\rm eV}$. 
The bias potential is fixed at $U_b = 0.3~{\rm eV}$. We find that as the 
couplings become more negative the frequency of oscillation of the 
transient current increases. The decrease in the value of the couplings
imply that the energy needed for the particle to hop from one site to a 
neighboring site is decreased. This frees up the particle, thereby allowing 
higher oscillation frequencies in the transient current. For long times after 
the transient oscillations have decayed away, the steady-state value of the 
current is independent of the specific value of the couplings.

Furthermore, we examine the effects of varying $v$ and $v^{\rm LR}$ 
separately. When $v^{\rm LR}$ is different from $v$, the interleads distance 
is different from the nearest-neighbor distance between sites in the leads. 
This results in a potential barrier that is different at the interleads 
coupling and thus, a particle moving from the left lead scatters at the 
interleads coupling. Figure \ref{fig:varyv}(b) shows the current as a 
function of time when the nearest-neighbor hopping parameter is set at 
$v = -2.7~{\rm eV}$ and we vary $v^{\rm LR}$ to values $-2.7~{\rm eV}$, 
$-2.4~{\rm eV}$, and $-2.1~{\rm eV}$. We do not consider 
$\left|v^{\rm LR}\right|>\left|v\right|$ values because that would imply a 
shorter interleads distance than the natural nearest-neighbor distance, 
represented by $v$, in the leads. In contrast, decreasing 
$\left|v^{\rm LR}\right|$ increases the potential barrier at the interleads 
coupling, and thus implying a longer interleads distance, and results in the 
reduction in the amplitude of the oscillating transient current. From 
Fig.~\ref{fig:varyv}(b), we also see that the peaks are slightly shifted to 
later times. In addition, the long-time steady-state current slightly 
decreases when $\left|v^{\rm LR}\right|$ is decreased. If we want to 
calculate the steady-state current using the Landauer formula, we would find 
that the transmission coefficient is reduced when $v^{\rm LR}$ is different 
from $v$ because of the scattering occuring at the interleads coupling. 

\begin{figure}[h!]
\includegraphics[width=3in,clip]{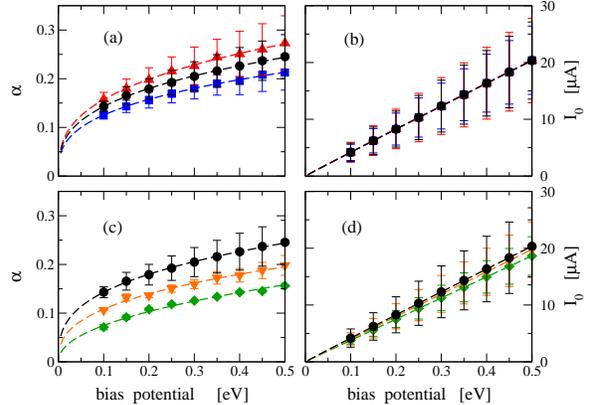}
\caption{(Color online) The exponent $\alpha$ and coefficient $I_0$ when $v$ 
and $v^{\rm LR}$ are varied. In (a) and (b), $v = v^{\rm LR} = -2.0~{\rm eV}$ 
for the (red triangles), $v = v^{\rm LR} = -2.7~{\rm eV}$ for the (black
dots), and $v = v^{\rm LR} = -4.0~{\rm eV}$ for (blue squares). In (c)
and (d), we set $v = -2.7~{\rm eV}$ and vary $v^{\rm LR}$ to values 
$-2.1~{\rm eV}$ (green diamonds), $-2.4~{\rm eV}$ (orange inverted 
triangles), and $-2.7~{\rm eV}$ (black dots). The dashed lines are the
power-law fits to the corresponding data points.
\label{fig:varyvLR}}
\end{figure}

The oscillating and decaying transient current when $v$ and $v^{\rm LR}$ 
are varied also follow a power law. In Figs.~\ref{fig:varyvLR}(a) and 
\ref{fig:varyvLR}(b) the values of $v$ and $v^{\rm LR}$ are varied together 
while in Figs.~\ref{fig:varyvLR}(c) and \ref{fig:varyvLR}(d) the values are 
varied separately. Figure \ref{fig:varyvLR}(b) shows that the values of $I_0$ 
is the same for the cases examined. Compared to Fig.~\ref{fig:varyvLR}(d), we 
see that the values of $I_0$ are slightly different. Identifying $I_0$ as the 
steady-state current, we thus confirm that it is the same whenever 
$v = v^{\rm LR}$. On the other hand, the scattering that happens at the 
interleads coupling when $v^{\rm LR}$ is different from $v$ affects the value 
of the steady-state current. Moreover, the plots of the exponent $\alpha$ and 
coefficient $I_0$ as functions of the bias potential can also be fitted to 
power laws. As shown in Table~\ref{tbl:fits}, when $v = v^{\rm LR}$, 
decreasing the value of the couplings decreases both the coefficient 
$\alpha_0$ and the exponent $\beta$, while $I_{00}$ and $\gamma$ remain the 
same.  The transient current therefore decays slower. This is because 
decreasing the couplings decreases the energy required for the particle to 
move around. This increase in the particle's freedom to move increases the 
frequency of oscillation and slightly lengthens the decay of the transient 
current. However, fixing the value of $v$ and increasing $v^{\rm LR}$ 
decreases $\alpha_0$ and $I_{00}$, but increases $\beta$. Therefore, for a 
given bias potential, increasing $v^{\rm LR}$ lengthens the decay but 
suppresses the amplitude of oscillation of the transient current. The 
parameters $\alpha_0$, $\beta$, $I_{00}$, and $\gamma$ depend on the type of 
material used. The value of the interleads coupling $v^{\rm LR}$, in addition, 
depends on the distance between the leads. The farther apart are the two 
leads, the higher is the value of $v^{\rm LR}$ because of the higher 
interleads potential barrier. Our results show that stronger interleads 
scattering lengthens the decay of the transient current. However, the 
scattering also suppresses the amplitude of the transient current and 
decreases the eventual value of the long-time steady-state current.

\begin{table}
\caption{(Color online) Values from the power-law fits to the exponent 
$\alpha$ as a function of the bias potential $U_b$ when the switch-on of the 
interleads coupling is in the form of a Heaviside step function. The shaded 
entries indicate cases when $v \neq v^{\rm LR}$. The dimension of $\alpha_0$ 
is $(1/{\rm eV})^{\beta}$ and $I_{00}$ is $\mu {\rm A}/({\rm eV})^{\gamma}$.}
\label{tbl:fits}
\begin{ruledtabular}
\begin{tabular}{|c|c|c|c|c|c|}
  \hline
  $v$ [eV] & $v^{\rm LR}$ [eV] & ~$\alpha_0$~ & ~$\beta$~ & $I_{00}$ 
  & $\gamma$ \\
  \hline \hline
  -2.0 & -2.0 & 0.344 & 0.340 & 40.125 & 0.980 \\
  -2.7 & -2.7 & 0.307 & 0.332 & 40.214 & 0.983 \\
  -4.0 & -4.0 & 0.268 & 0.330 & 40.305 & 0.985 \\
  \rowcolor[rgb]{1,1,0.4}
  -2.7 & -2.1 & 0.219 & 0.466 & 37.152 & 0.993 \\
  \rowcolor[rgb]{1,1,0.4} 
  -2.7 & -2.4 & 0.252 & 0.367 & 39.344 & 0.992 \\
  \hline
\end{tabular}
\end{ruledtabular}
\end{table}

\begin{figure}[h!]
\includegraphics[width=2.6in,clip]{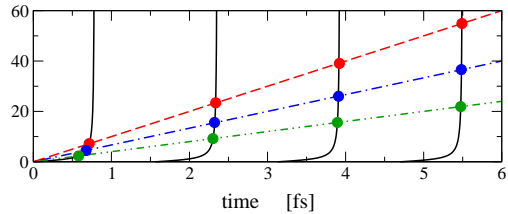}
\caption{(Color online) The location of the peaks in the transient current. 
The (black) lines represent $\tan(\omega~t + \phi)$, where 
$\omega = 2~{\rm [1/t]}$ and $\phi = \pi$. The (red) dashed line represents 
$\omega~t/\alpha$, where $\alpha = 0.2$. The (blue) dashed-dot line is when 
$\alpha = 0.3$ and the (green) dashed-double dot line is when $\alpha = 0.5$. 
The times when the straight and tangent lines intersect correspond to the 
times when the peaks in the transient current occur.
\label{fig:peaks}}
\end{figure}

The times when the peaks in the current occur can be known from the 
extremum of the power-law form of the current in Eq.~(\ref{eq:powerlaw}). 
Taking the time derivative of Eq.~(\ref{eq:powerlaw}) and then equating the 
result to $0$, we find the extremum of the current to occur at times $t_p$ 
whenever the following is satisfied:
\begin{equation}
\frac{\omega~t_p}{\alpha} = \tan\left(\omega~t_p + \phi\right).
\label{eq:tp}
\end{equation}
The left-hand side is an equation for a straight line with a slope that 
depends on $\alpha$. Since $\alpha$ varies depending on the values of the
bias potential, the couplings, and the speed of the switch-on, changing these
parameters would change the slope. As a consequence, the location in time of 
the current peaks would also change. This can be seen by noting how the peaks 
in the transient current move in Fig.~\ref{fig:tanh} as $\alpha$ is varied. 
$t_p$ can be determined by the intersection points of the straight and tangent 
lines, corresponding to the left-hand side and right-hand side, respectively, 
of Eq.~(\ref{eq:tp}) and as shown in Fig.~\ref{fig:peaks}. The times when the 
current peaks occur are located whenever the two curves intersect. Since the 
slope of the straight line depends on $\alpha$, we see that the faster 
decaying transient current, i.e., higher values of $\alpha$, correspond to 
earlier peak times.

\begin{figure}[h!]
\includegraphics[width=3in,clip]{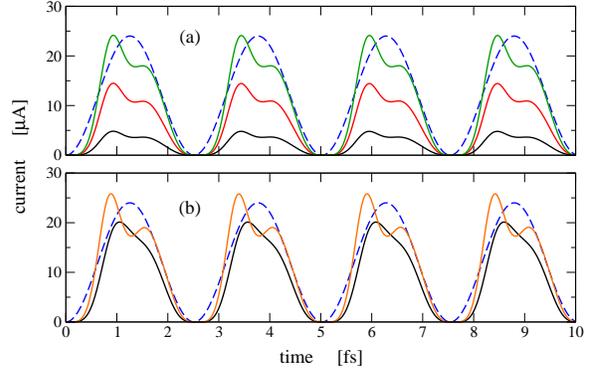}
\caption{(Color online) The current as a function of time for the 
nano-oscillator with driving frequency $\omega_d = 0.25~{\rm [1/t]}$. (a) The 
bias potentials are $U_b = 0.1~{\rm eV}$ (black line), $U_b = 0.3~{\rm eV}$ 
(red line), and $U_b = 0.5~{\rm eV}$ (green line). The couplings are
$v = v^{\rm LR} = -2.7~{\rm eV}$. (b) The bias potential is 
$U_b = 0.5~{\rm eV}$ and the hopping parameter is $v = -2.7~{\rm eV}$.
The interleads coupling has amplitudes $v^{\rm LR} = -2.1~{\rm eV}$ 
(black line) and $v^{\rm LR} = -3.0~{\rm eV}$ (orange line). The (blue) 
dashed lines show the harmonic variation of the interleads coupling.
Their amplitudes are not drawn to scale.
\label{fig:ncos}}
\end{figure}

Finally, we investigate the transport properties of the device having a 
regular time variation, such as a nano-oscillator. In a nano-oscillator, the 
rotating disk in Fig.~\ref{fig:main}(a) is rocked back and forth across the 
dashed line. This would result in a harmonic modulation of the interleads 
coupling and would dynamically modulate the current through the device. 
However, compared to an alternating current which changes sign, the modulated 
current flowing out of the nano-oscillator maintains the same sign. We model 
the harmonically modulated coupling in the form 
$v^{\rm LR}(t) = v^{\rm LR}/2 \cdot \left(1 - \cos \omega_d t\right)$,
where $\omega_d$ is the driving frequency of modulation. Figure \ref{fig:ncos} 
shows the current characteristics as a function of time as the interleads 
coupling is swinged back and forth with driving frequency 
$\omega_d = 0.25~{\rm [1/t]}$. In Fig.~\ref{fig:ncos}(a), the couplings are 
$v = v^{\rm LR} = -2.7~{\rm eV}$ and the bias potential is varied. In
Fig.~\ref{fig:ncos}(b) the bias potential is set at $U_b = 0.5~{\rm eV}$, the 
hopping parameter is fixed at $v = -2.7~{\rm eV}$, and the interleads coupling 
$v^{\rm LR}$ is varied. We find that the current through the oscillator comes 
in pulses. However, it does not exactly follow the harmonic form of the 
coupling. The interleads coupling is maximum at times when the left and right 
leads are exactly aligned. In contrast, we see that the peaks in the current 
do not coincide with the times when the interleads coupling is maximum. The 
shape of the curve for the current actually looks like the truncated version 
of the transient current we examined in Fig.~\ref{fig:tanh}. In particular, 
the initial overshoot of the transient current manifests as the first current 
peak in Fig.~\ref{fig:ncos}. This peak, however, does not occur when the 
interleads coupling is maximum. In addition, the times when the peak occurs 
depend on the values of the bias potential and the couplings. This dependence 
of the peak location to the above physical parameters follow similar 
dependence of the peak location in the nano-relay. In the design of 
nano-circuits containing an oscillator, therefore, it should be noted 
that the maximum current does not occur when the leads are exactly aligned and 
that the exact location of these peaks depend on the values of the applied 
bias potential, the nearest-neighbor coupling, and the interleads coupling.

%----------------------------------------------------------------------------
%                        Summary and conclusion
%----------------------------------------------------------------------------
\section{SUMMARY AND CONCLUSION}
\label{sec:summary}

In summary, we examine a device that could act as a nano-relay or a 
nano-oscillator. The device consists of two leads and a time-varying 
interleads coupling. We use NEGF to derive a nonperturbative expression for 
the time-dependent current flowing from one lead to the other.  In the 
nano-relay configuration, we model the switch-on of the interleads coupling 
in the form of either a step function or a slowly progressing hyperbolic 
tangent. We find that the current oscillates and decays in time just after 
switch-on and during the transient regime. In both the step function and 
hyperbolic tangent switch-on, the decay of the transient current fits a power 
law. This leads to an equivalent RLC series circuit where all of the 
components have dynamical properties. We also find that the values of the 
couplings $v$ and $v^{\rm LR}$, the scattering at the interleads coupling, 
the speed of the switch-on, and the value of the bias potential affect the 
decay time of the transient current. In the long-time regime, the current 
approaches the steady-state value. In the nano-oscillator, we model the 
dynamical sytem by harmonically modulating the interleads coupling. We find 
that the current passes through the device in pulses, maintains the same sign, 
but does not exactly follow the functional form of the oscillating coupling. 
In particular, the peaks in the current do not occur at the times whenever the 
leads are exactly aligned.

The expressions for the current shown in Eq.~(\ref{eq:rightcurrent}) and the 
corresponding lesser Green's function shown in Eq.~(\ref{eq:gless}) are 
general and should be applicable to transport in quasi-linear systems where a 
switch-on in time occurs. The current oscillates and decays as a power law
after a switch-on. This power-law decay implies the presence of dynamical 
resistance, inductance, and capacitance components.

%----------------------------------------------------------------------------
%                        Acknowledgment
%----------------------------------------------------------------------------
\begin{acknowledgments}

We would like to acknowledge Jian-Sheng Wang, Vincent Lee, and Kai-Tak Lam
for insightful discussions. This work is supported by A*STAR and SERC under 
Grant No. 082-101-0023. Computational resources are provided by the 
Computational Nanoelectronics and Nano-Device Laboratory, Department of 
Electrical and Computer Engineering, National University of Singapore.

\end{acknowledgments}

%----------------------------------------------------------------------------
%                             Bibliography
%----------------------------------------------------------------------------


\begin{thebibliography}{99}
\bibitem{dc}
  It is relevant to notice in Fig.~\ref{fig:tanh} that the current approaches
  a non-zero steady-state value during later times. This indicates the 
  presence of a dc component in the current and is due to the applied static 
  bias voltage $V_b$ between the left and right leads. The dc component 
  vanishes when there is no applied bias voltage. If a dc component in the 
  current is desired without the application of a bias voltage between the 
  leads (and assuming that the leads have the same chemical potentials), then 
  it is necessary to break time-reversal and spatial-inversion symmetry in the 
  system.\cite{arrachea05b}
\end{thebibliography}
\end{document}